\documentclass[10pt]{wlscirep}

\usepackage{bm}
\usepackage{amssymb}
\usepackage{amsmath}
\usepackage{braket}
\usepackage{graphicx}
 \usepackage{color}

\title{
Theory for spin torque in Weyl semimetal with magnetic texture
}

\author[1,2,*]{Daichi Kurebayashi}
\author[2,3]{Kentaro Nomura}
\affil[1]{Center for Emergent Matter Science, RIKEN, Wako 351-0198, Japan}
\affil[2]{Institute for Materials Research, Tohoku University, Sendai 980-8577, Japan}
\affil[3]{Center for Spintronics Research Network, Tohoku University, Sendai 980-8577, Japan}
\affil[*]{daichi.kurebayashi@riken.jp}

\begin{abstract}
The spin-transfer torque is a fundamental physical quantity to operate the spintronics devices such as racetrack memory.
We theoretically study the spin-transfer torque and analyze the dynamics of the magnetic domain walls in magnetic Weyl semimetals.
Owing to the strong spin-orbit coupling in Weyl semimetals, the spin-transfer torque can be significantly enhanced, because of which they can provide a more efficient means of controlling magnetic textures.
We derive the analytical expression of the spin-transfer torque and find that the velocity of the domain wall is one order of magnitude greater than that of conventional ferromagnetic metals.
Furthermore, due to the suppression of longitudinal conductivity in the thin domain-wall configuration, the dissipation due to Joule heating for the spin-transfer torque becomes much smaller than that in bulk metallic ferromagnets.
Consequently, the fast-control of the domain wall can be achieved with smaller dissipation from Joule heating in the Weyl semimetals as required for application to low-energy-consumption spintronics devices.
\end{abstract}

\begin{document}

\flushbottom
\maketitle

\thispagestyle{empty}

While microfabrication technology in electronic devices is reaching the limits of the atomic scale, the idea of utilizing an electron's spin degrees of freedom, in addition to the electronic degrees of freedom, to further improve the performance of devices has garnered significant attention in spintronics in both academia and the industry \cite{Zutic2004}.
Controlling the dynamics of magnetization is a challenge for the successful application of spintronic memory, logic, and sensing devices.
Local magnetic fields have been used for this purpose, but applying local magnetic fields causes difficulties in creating scalable systems.
Instead, current-induced spin torque, such as the spin-transfer torque, has been widely used for the electrical manipulation of magnetization \cite{Slonczewski1996,Berger1996,Ralph2008,Brataas2012}.
The magnetic racetrack memory, for instance, has been proposed as the promising application of current-induced spin torque to spintronics devices \cite{Parkin2008}.
The sizeable current density required to operate the device, however, limits its efficiency because of Joule heating, thus impeding its commercial application.
Hence, more efficient means of controlling magnetization are indispensable to the application.

The application of topological properties has attracted research interest to more efficiently manipulate magnetization.
For instance, at the interface of the topological insulator and the ferromagnetic insulator \cite{Hasan2010,Qi2011}, the electrical control of magnetic textures \cite{Nomura2010,Hurst2015}, magnetization switching induced by electric current \cite{Garate2010,Yokoyama2010,Sakai2014,Fan2014}, and spin-charge conversion \cite{Shiomi2014,Kondou2016,Okuma2016} have been studied theoretically and experimentally.

Recently, Weyl semimetals \cite{ShuichiMurakami2007,Wan2011,Burkov2011} have emerged as a new class of topological material characterized by gapless points in the bulk---the so-called Weyl nodes---and the breaking of the inversion or time-reversal symmetry.
Close to the Weyl nodes, excitation can be described by a three-dimensional (3D) linear dispersion which is an analog of the Weyl fermion in high-energy physics \cite{Weyl1929}.
Due to this relativistic electronic structure, new physics beyond the topological insulator, such as the chiral anomaly, can be expected \cite{Nielsen1981,Hosur2013}.
In particular, the Weyl semimetal realized by spontaneous ferromagnetism \cite{Bulmash2014,Kurebayashi2014} possesses both topological and magnetic properties, and thus may be a promising candidate for application to spintronics devices such as a race-track memory.
Such a magnetic Weyl semimetal phase was recently observed in $\rm Co_3Sn_2S_2$ \cite{Xu2017a,Liu2018}, which showed a large anomalous Hall angle indicating the strong spin-orbit coupling and small longitudinal conductivity.
This new experimental finding has the potential to further accelerate development in topological spintronics realizing low-energy consumption electrically-induced spin torques.

It was proposed that magnetic textures in Weyl semimetals retain a localized electric charge\cite{Araki2016a,Araki2018a}, and that more efficient electrical manipulation of the magnetic texture is possible in Weyl semimetals than in conventional ferromagnetic metals owing to the strong correlation between magnetic texture and charge degrees of freedom.
The dynamics of magnetic textures are driven by electrically-induced spin torques which consist of adiabatic and non-adiabatic spin-transfer torques in conventional ferromagnetic metals.
Adiabatic spin-transfer torque results from the transfer of angular momentum while the non-adiabatic term is generated by spin-orbit coupling.
Because of the strong spin-orbit coupling in Weyl semimetals, the effect of non-adiabatic spin-transfer torque is expected to be larger, which also enhances the dynamics of magnetic textures.
However, quantitative studies on the effects of spin-transfer torque on systems with strong spin-orbit coupling, such as Weyl semimetals, have not been conducted, and are needed for the successful application of spintronics devices such as the race-track memory.
In this paper, we first study the electrically-induced spin torque exerted on magnetic textures in Weyl semimetals, and obtain the analytical expression of spin torque corresponding to non-adiabatic spin-transfer torque.

The remainder of this paper is organized as follows: We consider a continuum model for the ideal case where the spin-orbit interaction is isotropic, $\bm \sigma \cdot \bm k$, and referred to as the Weyl-type spin-orbit coupling (SOC), and analytically derive the spin torque acting on inhomogeneous local magnetization by computing the non-equilibrium spin density induced by the applied electric fields.
We then employ lattice models of Weyl semimetals with several types of spin-orbit coupling and numerically calculate the induced spin density.
Finally, we argue the dynamics of the domain wall driven by the obtained spin torque.
We estimate the velocity of a domain wall for the case where $\rm Co_3Sn_2S_2$, which is one order of magnitude faster than that in conventional ferromagnetic metals.

\section*{Results}
\subsection*{Formalism}
Magnetization dynamics can be described by the famous Landau—Lifshitz--Gilbert (LLG) equation:
\begin{eqnarray}
\frac{d\hat{\bm M}}{dt} = \gamma_0\bm B\times \hat{\bm M}+\alpha\hat{\bm M}\times\frac{d\hat{\bm M}}{dt}+\bm T_e,
\label{llg}
\end{eqnarray}
where $\hat{\bm M}(\bm r)=(M_x,M_y,M_z)$ is a normalized directional vector of local magnetic moments at $\bm r$, $\gamma_0$ is the gyromagnetic ratio, $\bm B$ is an effective field containing an external magnetic field, and $\alpha$ is the damping constant. The torque given by the first term describes magnetization precession around the effective magnetic field $\bm B$. The second term, called the Gilbert damping term, represents a torque which drives the magnetization vector in the direction of the effective field $\bm B$. However, once they have been placed in a material, an additional coupling between the magnetic moments and the electric fields arises through exchange interaction between the conduction electrons and the local moments. The effect of the background conduction electrons is known as spin torque, the last term of LLG equation, and is described by
\begin{eqnarray}
\bm T_e(\bm r) = \frac{JS}{\hbar\rho_S}\hat{\bm M}\times \langle\hat{\bm \sigma}(\bm r)\rangle
\label{torque_def}
\end{eqnarray}
where $J$ is the exchange coupling constant between the conduction electrons and the local moments, $S$ is the amplitude of the localized spin of magnetic moment, $\rho_S $ is the number of local magnetic elements per unit volume, and $\langle \bm \sigma(\bm r)\rangle$ is a non-equilibrium spin polarisation of itinerant electrons \cite{Zhang2004}. The electrically-induced spin torques, therefore, are obtained by calculating the non-equilibrium spin density of the electrons as a response to the external electric fields. Our interest in this paper is in the spin torques which arise when the magnetic configuration is spatially inhomogeneous. The electrically-induced spin density in momentum space can be expanded as
\begin{eqnarray}
 \langle \sigma_i(\omega,\bm q)\rangle = \chi^{(0)}_{ij}(\omega,\bm q)M_j(\omega,\bm q) + \chi^{(1)}_{ij}(\omega,\bm q)E_j +\chi^{(2)}_{ijk} (\omega,\bm q)M_j(\omega,\bm q) E_k+\chi^{(3)}_{ijkl} (\omega,\bm q)M_j(\omega,\bm q)M_k(\omega,\bm q) E_l
\end{eqnarray}
with respect to the external electric field, $E_i$, and magnetization $M_i$ \cite{Kohno2006,Hals2015}. The first term contributes as renormalization to the Gilbert damping and the magnitude of the spin, while the second term corresponds to the spin-orbit torque which vanishes in the presence of inversion symmetry. The third and fourth terms, respectively, correspond to non-adiabatic and adiabatic spin-transfer torques, which are the focus of this study. In case of the conventional Schr\"odinger electrons, the spin-transfer torques are known to be proportional to spin current $\bm j_s$, described by $\bm T_{\rm STT}\propto (\bm j_s\cdot\bm \nabla)\hat{\bm M}+\beta\hat{\bm M}\times(\bm j_s\cdot \bm \nabla)\hat{\bm M}$, where the first term is the adiabatic spin-transfer torque and the second is the non-adiabatic spin-transfer torque characterized by the dimensionless coefficient $\beta$. Usually, in conventional metals, non-adiabatic contribution is considerably smaller than adiabatic contribution, i.e. $\beta \sim 0.01$ \cite{Eltschka2010}. In strongly spin-orbit coupled systems, however, the analytical expression of spin-transfer torques can be drastically modified, e.g. the conventional spin-transfer torque vanishes at the surface of the magnetic topological insulator \cite{Yokoyama2010,Sakai2014}. As Weyl semimetals are usually realized in strongly spin-orbit coupled systems, the spin-transfer torque needs to be studied with a non-perturbative treatment of spin-orbit coupling.

As a model describing the electronic system, we consider 3D Weyl electrons coupled with the local moments of magnetic elements via exchange interaction. The low-energy electronic structure is described by the Hamiltonian \cite{Burkov2011,Zyuzin2012a},
\begin{eqnarray} 
H_{\rm WSM} &=& \int d^3r \psi^\dagger(\bm r)\left[-i\hbar v_F\tau_z\bm \sigma \cdot \bm \nabla -JS\hat{\bm M}(\bm r)\cdot \bm \sigma\right]\psi(\bm r).
\label{H0}
\end{eqnarray}
whose eigenvalues with a uniform magnetization are given by
$
\epsilon_{\lambda\pm}(\bm k) = \pm \hbar v_F\left|\bm k-\lambda JS\hat{\bm M}/(\hbar v_F)\right|.
$
The first term of the Hamiltonian describes a degenerating 3D massless Dirac electron, where $v_F$ is the Fermi velocity, $\bm \sigma=(\sigma_x,\sigma_y,\sigma_z)$ is the Pauli matrix representing real spin operators, and $\tau_z $ is the chirality operator, the eigenvalues of which are $\lambda = \pm 1$, labeling two Weyl nodes. Note that in materials in practice, the structure of the spin-orbit coupling is determined by crystal symmetry. Other types of spin-orbit coupling are numerically examined in the second half of the paper. The second term is the exchange coupling between Weyl electrons and the localized magnetic moments, characterizing the time-reversal-symmetry violation of the system. The band structure of the Hamiltonian composed of two Weyl nodes separated in momentum space is given by $2JS\hat{\bm M}/(\hbar v_F)$.

It is worth noting that the Hamiltonian, Eq. (\ref{H0}), commutes with the chirality operator, $\tau_z$. As with the spin currents, $\bm j_S=\bm j_\uparrow-\bm j_\downarrow$, we can naturally introduce axial currents, the difference between currents with positive and negative chirality, namely
\begin{eqnarray}
\langle \bm j^5\rangle \equiv\langle \bm j_+\rangle - \langle\bm j_-\rangle = -e\Braket{\tau_z\frac{\partial H_{\rm WSM}}{\hbar \partial \bm k}}= -ev_F \langle \bm \sigma\rangle
\label{correspondence}
\end{eqnarray}
which coincides with the expectation value of spin except for the proportional constant. This suggests that the Hamiltonian, Eq. (\ref{H0}), possesses a one-to-one correspondence between the axial currents and the spin. The electron's spin couples to the local moments via exchange interaction as $-JS\bm M\cdot \bm \sigma$, while the axial currents couple to external fields as $-e\bm A_5\cdot\bm j^5$, where $\bm A_5$ is axial vector potential that changes sign depending on the chirality of the electrons. Due to the correspondence between spin and the axial current operators, the electrons cannot distinguish between these perturbations, namely the magnetization of local moments and the axial vector potential coupled to the electrons in the same way. Furthermore, due to the correspondence between the magnetization and the axial vector potential, the curl of the magnetization acts as the emergent magnetic field as
\begin{eqnarray}
\bm B_5 = \bm \nabla \times \bm A_5 = \frac{JS}{ev_F}\bm \nabla \times \hat{\bm M}.
\end{eqnarray}
which changes sign depending on the chirality of the Weyl fermions, the so-called the axial magnetic field \cite{Araki2018}. Based on the correspondence in our model, the spin density induced under magnetic texture can be calculated by evaluating the axial current density under axial magnetic fields.


\subsection*{Spin torque}
The axial current density with an inhomogeneous magnetic configuration is calculated as 
$
\langle  j^{5}_i(\bm r) \rangle = \sigma_{ij}(\bm B_5) E_j
$
where $\sigma_{ij}$ is the conductivity tensor. In the semiclassical regime, the Hall component of the conductivity tensor is given by
\begin{eqnarray}
 \sigma_{ij}(\bm B_5) =\frac{e^3v_F}{3\hbar^3\pi^2} \frac{\tau^2E_F}{1+(\tau\omega_C)^2}\epsilon_{ijk}B_5^k= \frac{e^2JS}{3\hbar^3\pi^2} \frac{\tau^2E_F}{1+(\tau\omega_C)^2}\epsilon_{ijk}(\bm \nabla \times \hat{\bm M})_k
\end{eqnarray}
where $\omega_C= \left|\frac{ev_F^2\bm B_5(\bm r)}{E_F}\right|=\left|\frac{JSv_F\bm \nabla\times \hat{\bm M}}{E_F}\right|$ is cyclotron frequency owing to the curl of magnetization, $\tau$ is the relaxation time, and $E_F$ is Fermi energy measured from the Weyl points. For the derivation of conductivity, see the Methods section. In axial transport, the coefficient corresponds to the ordinary Hall conductivity for axial currents caused by axial magnetic fields.
With the slowly-varying magnetization, namely $\omega_C\tau <<1$, the Hall conductivity is calculated as
$
\nonumber \sigma_{ij}(\bm M)= \frac{e^2JS\tau^2E_F}{3\hbar^3\pi^2}\epsilon_{ijk}(\bm \nabla \times \hat{\bm M})_k
$
which is in the second order of the relaxation time.
As the spatial variation of magnetization becomes steep, the dependence of relaxation time to the Hall conductivity gets weaker.
When spatial variation of magnetization forms Landau levels, corresponding to $\omega_C\tau >>1$, the system eventually turns into the quantum Hall regime.
In this regime, the semiclassical analysis is no longer applicable.
Instead, the Hall conductivity for the 3D Dirac electrons in the quantum regime has been well studied
\cite{Araki2018}, and is given by
\begin{eqnarray}
\sigma_{ij}(\bm M)=\frac{e^2E_F}{2\hbar^2\pi^2v_F}\epsilon_{ijk}\frac{(\bm \nabla \times \hat{\bm M})_k}{|\bm \nabla \times \hat{\bm M}|}
\end{eqnarray}
becoming independent of field strength.

Finally, we can calculate the induced-spin density by the relation between spin and the axial current, Eq. (\ref{correspondence}), as
\begin{eqnarray}
 \langle \bm \sigma(\bm r) \rangle = \chi_{\rm S}\left[\bm \nabla \times \hat{\bm M}(\bm r)\right]\times \bm E
\label{result}
\end{eqnarray}
where spin susceptibility is defined by
\begin{eqnarray}
\chi_{\rm S}\equiv
\begin{cases}
\frac{eE_F}{2\hbar^2\pi^2v_F^2}\epsilon_{ijk}\frac{(\bm \nabla \times \hat{\bm M})_k}{|\bm \nabla \times \hat{\bm M}|}&\left(\mbox{Semiclassical regime}\right)\\
\frac{eE_F}{2\hbar^2\pi^2v_F^2 |\bm \nabla \times \bm{\hat{M}}|}&\left(\mbox{Quantum Hall regime}\right)
\end{cases}.
\end{eqnarray}
Substituting Eq. (\ref{result}) into Eq. (\ref{torque_def}), the associated spin torque is then given by 
\begin{eqnarray}
\bm T_e (\bm r) = \frac{JS}{\hbar \rho_S}\chi_S\hat{\bm M}(\bm r)\times\left[ \left(\bm \nabla \times \hat{\bm M}(\bm r)\right)\times \bm E\right]=\frac{JS}{\hbar \rho_S}\chi_S\left[\hat{\bm M}(\bm r)\times (\bm E\cdot \bm \nabla)\hat{\bm M}(\bm r)-\hat{\bm M}(\bm r)\times \bm \nabla \left\{\hat{\bm M}(\bm r)\cdot \bm E\right\}\right].
\label{spin torque}
\end{eqnarray}
The result shows that the spin torque is induced when the magnetization spatially varies in the presence of an external electric field, which is similar to the spin-transfer torque in conventional metals. Indeed, the first term in the final expression of Eq. (\ref{spin torque}) coincides with the non-adiabatic spin-transfer torque. 
In slowly-varying magnetization regime where $\omega_C\tau<<1$, the spin-transfer torque appears as the second order of relaxation time $\tau$ in Weyl semimetals, whereas it is the first order in conventional ferromagnetic metals \cite{Kohno2006}. 
On the contrary, in the quantum Hall regime where $\omega_C\tau>>1$, this susceptibility becomes independent of relaxation time, indicating that the spin torque is an intrinsic phenomenon. Furthermore, as the spatial variation of magnetization becomes steeper, the longitudinal electric conductivity asymptotically reaches zero (see the Methods). This suggests that the spin torque is driven by an electric field rather than electric currents. Therefore, dissipation from Joule heating is significantly suppressed and becomes much smaller than that in conventional ferromagnetic metals.
 
In past work, a localized charge at the domain wall in magnetic Weyl semimetals has been theoretically proposed \cite{Araki2016a,Araki2018a}. The origin of the localized charge is understood as the enhanced density of states by Landau-level degeneracy owing to the axial magnetic field. These results suggest that this localized charge attached to the domain wall might allow us to manipulate magnetic texture by external electric fields. 
Our result reveals that the spin-torque, indeed, is induced due to the magnetic texture in the Weyl semimetals and gives a physical understanding of the effect as the axial Hall effect.

In this subsection, we have derived the analytical expression of the spin torque while naively assuming the axial magnetic field generated by the curl of magnetization is uniform in space.
Our analytical result is still applicable for slowly-varying magnetization where the axial magnetic field is considered as locally uniform.
However, if we consider steep magnetic domain walls, for instance, a spatial profile of the axial magnetic field is strongly inhomogeneous.
Therefore, we need further analysis beyond the semiclassical approach to understand the dynamics of steep magnetic textures.
Also, we only consider the Weyl-type SOC, $\bm \sigma \cdot \bm k$, where the Weyl semimetal phase is realized for any direction of magnetization.
This is, however, not always the case for materials in practice.
Our derivation is based on the correspondence between the spin and axial currents, which is applicable only to the Weyl-type SOC.
Thus, it is important to examine the induced spin density for the steep magnetic texture with other types of spin-orbit couplings.
\subsection*{Numerical study}
In the following,  we employ lattice models describing a Weyl semimetal with three types of spin-orbit couplings and numerically compute the induced spin density for the steep magnetic texure.
As spin-orbit coupling, we consider three cases: Weyl-type SOC, Rashba-type SOC, and $s_z$-conserved SOC. We assume a four-band Dirac Hamiltonian with two-fold degeneracy in the presence of time-reversal symmetry in the slab geometry stacking along the $x$-direction. The effective Hamiltonian on the cubic lattice is given as
\begin{eqnarray}
\mathcal{H}_{\rm Dirac} = \sum_{n_x \bm p}c_{n_x \bm p}^\dagger H_\parallel(\bm p)c_{n_x \bm p}+\frac{1}{2}\sum_{n_x \bm p}\left[c_{n_x \bm p}^\dagger T_\perp c_{n_x+1 \bm p}+\mbox{h.c}\right]
\end{eqnarray}
where $c_{n_x \bm p}=\left[c_{n_x \bm p\uparrow-},c_{n_x \bm p\downarrow-},c_{n_x \bm p\uparrow+},c_{n_x \bm p\downarrow+}\right]^T$ is an electron annihilation operator, $\pm$ and $\uparrow,\downarrow$ denote the orbital and spin degrees of freedom, respectively, and $\bm p = (p_y,p_z)$ is an in-plane momentum. For the Weyl-type SOC and Rashba-type SOC, the in-plane and out-of-plane hopping matrices of the Hamiltonian are given by
$
H_\parallel(\bm p) = t\left[\sin(p_y) \alpha_2 + \sin(p_z) \alpha_3\right] + r\left[3- \cos(p_y)-\cos(p_z)\right]\alpha_4
$
, and
$
T_\perp = -(it\alpha_1 + r\alpha_4)/2,
$
respectively \cite{Zhang2009,Liu2010}. Dirac's alpha matrices are defined as
\begin{eqnarray}
\nonumber \mbox{(Weyl-type SOC)}\hspace{3.5cm}&&\hspace{1cm}\mbox{(Rashba-type SOC)}\\
\nonumber\alpha_1 = \tau_x\sigma_x,
\alpha_2 = \tau_x\sigma_y,
\alpha_3 = \tau_x\sigma_z,
\alpha_4 = \tau_z\sigma_0
&&\hspace{1cm}\alpha_1 = \tau_x\sigma_y,
\alpha_2 = -\tau_x\sigma_x,
\alpha_3 = \tau_y\sigma_0,
\alpha_4 = \tau_z\sigma_0
\end{eqnarray}
where $\sigma_i$ and $\tau_i$ represent the spin and the orbital degrees of freedom, respectively. The Hamiltonian for the $s_z$-conserved SOC\cite{Wang2012,Morimoto2014} is given by
$
H_\parallel(\bm p) = -t\sin(p_y)\tau_y\sigma_0+r\left[2-\cos(p_y)-\cos(p_z)\right]\tau_z\sigma_0
$
and
$
T_\perp = -\left(i t \tau_x\sigma_z+ r \tau_z\sigma_0\right)/2.
$
The Hamiltonian coincides with Eq. (\ref{H0}) in the continuum limits for the Weyl-type SOC, whereas the Rashba-type SOC describes the electronic structure of layered materials, such as $\rm Bi_2Se_3$ \cite{Zhang2009,Liu2010}. Both models exhibit degenerated Dirac cones at the $\Gamma$ point when time-reversal symmetry is present. The model with the $s_z$-conserved SOC, on the contrary, was proposed to describe a topological Dirac semimetal phase \cite{Wang2012,Wang2013a,Liu2014,Liu2014b} possessing two degenerate Dirac cones at $p_z = \pm\pi/2$. Note that the $z$-component of spin is conserved for the $s_z$-conserved SOC. It has been recently proposed that an effective model for the ferromagnetic Weyl semimetal, $\rm Co_3Sn_2S_2$ \cite{Xu2017a,Liu2018}, can be described by a model with the $s_z$-conserved SOC and exchange interaction \cite{Ozawa2018}. When the exchange energy is larger than the bandwidth, two copies of the degenerate bands split in terms of energy, and the system moves to the half-metallic Weyl semimetal phase.

For interaction between itinerant electrons and the localized moments, we introduce both orbital-independent and -dependent exchange interactions:
\begin{eqnarray}
\mathcal{H}_{\rm ex} = \sum_{n_x\bm p} c_{n_x\bm p}^\dagger\left[J_0\hat{\bm M}(n_x)\cdot \bm\sigma+J_1\tau_z\hat{\bm M}(n_x)\cdot \bm\sigma\right]c_{n_x\bm p}
\end{eqnarray}
where $J_0$ and $J_1$ are the orbital-independent and -dependent exchange constants, respectively. For the Weyl-type SOC, two Weyl nodes appear along the momentum axis parallel to magnetization when $J_0>J_1$, whereas there appears a line node in a plane perpendicular to the magnetization when $J_0<J_1$ \cite{Burkov2018}. On the contrary, for the Rashba-type SOC with $J_0>J_1$, the Weyl nodes appear only along the $k_z$-axis, and the line node appears when the magnetization is perpendicular to the $z$-axis.

\begin{figure}[htbp]
\centering
\includegraphics[width=0.8\linewidth]{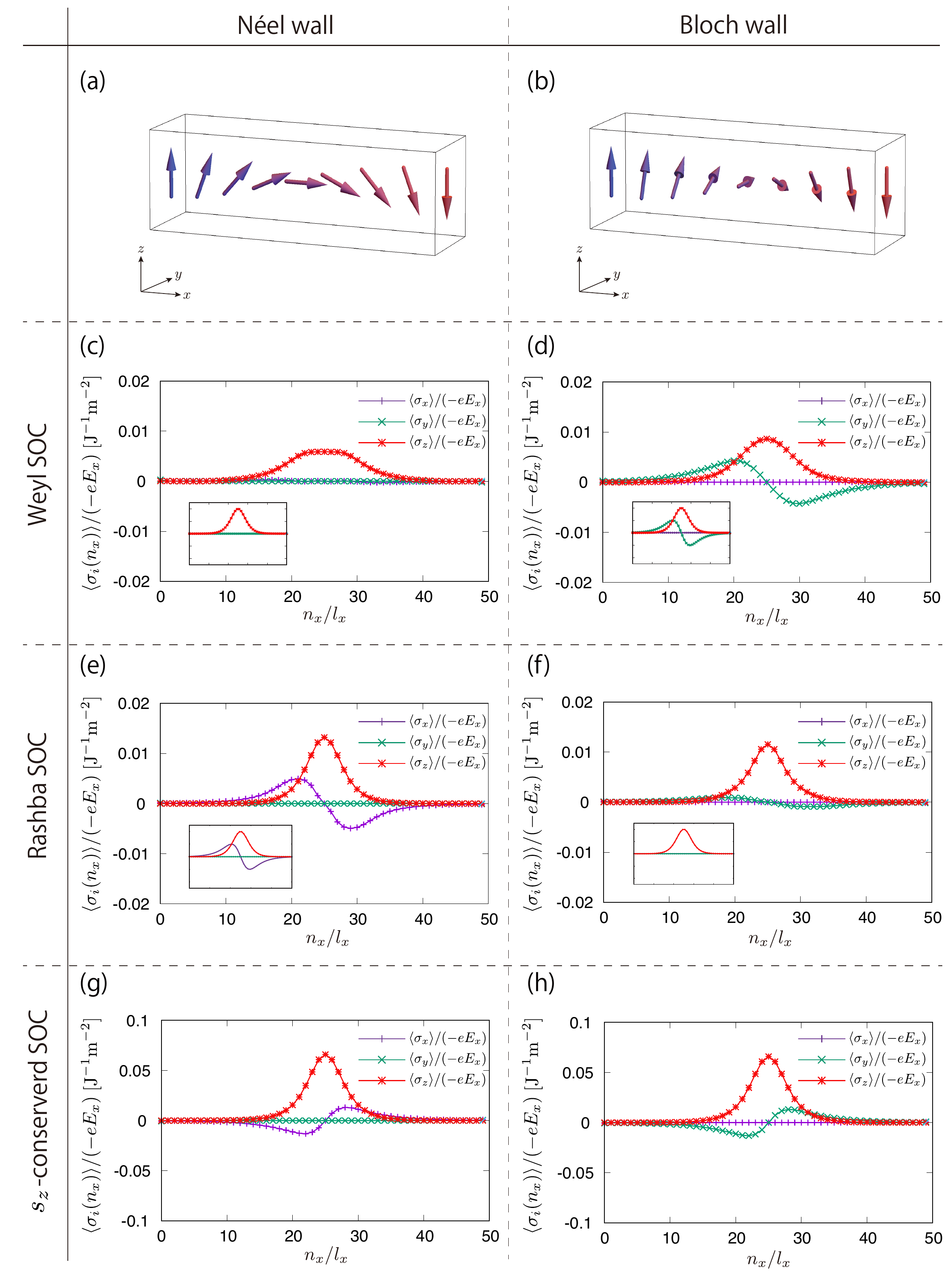}
\caption{
(a), (b) Magnetic configuration.
The magnetic structures are given by
$
\hat{\bm M} (n_x)=1/\cosh\left(l_x n_x/\xi\right)\bm n+\tanh\left(l_x n_x/\xi\right)\hat{\bm z},
$
where $\bm n = \hat{\bm x}$ for the N\'eel wall and $\bm n = \hat{\bm y}$ for the Bloch wall, $l_x$ is a lattice constant, and $\xi$ is the width of the domain wall taken as $\xi/l_x = 5$. (c), (d) Spin density of the Weyl-type SOC under the N\'eel walls and the Bloch wall, respectively. (e), (f) Spin density of the Rashba-type SOC under the N\'eel walls and the Bloch wall, respectively. Insets show the analytical results expected from the continuum models. (g), (h) Spin density of the $s_z$-conserved SOC under the N\'eel walls and the Bloch wall, respectively. The Hamiltonian parameters were chosen as $t = r= 1$ and $E_F/t = 0.1$. The exchange coupling constants were $J_0/t = 1$ and $J_1/J_0 = 0.2$ for the Weyl-type SOC and the Rashba-type SOC, respectively, and $J_0/t = 3$ and $J_1/J_0 = 0.2$ for the $s_z$-conserved SOC. 
For the $s_z$-conserved SOC, we used a lager exchange coupling constant to have half-metallic electronic structure.
}
\label{fig1}
\end{figure}

For the numerical simulation, we considered a N\'eel and a Bloch domain wall as the simplest magnetic textures with finite rotations. The spin density induced by the electric field applied along the $x$-direction was calculated as
\begin{eqnarray}
\frac{\langle\sigma_i (n_x)\rangle}{-e E_x}& = &\sum_{\bm p}\sum_{n,m}\frac{
\bra{n,\bm p}v_x\ket{m,\bm p}\bra{m,\bm p}\sigma_i(n_x)\ket{n,\bm p}
}{(\epsilon_{n\bm p}-\epsilon_{m\bm p})^2}\left[f(\epsilon_{n\bm p})-f(\epsilon_{m\bm p})\right]
\end{eqnarray}
where $\epsilon_{n \bm p}$ and $\ket{n,\bm p}$ are the $n$th eigenvalue and eigenvector, and $v_x$ and $\sigma_i(n_x)$ are the velocity and spin operator, respectively. We numerically diagonalize the lattice Hamiltonian under the N\'eel and Bloch domain walls shown in FIGs. \ref{fig1} (a) and (b) to obtain the eigenvalues and eigenvectors, and calculate the spin density, shown in FIG.\ref{fig1}. The results clearly show that spin density was locally induced at a domain wall. Figures \ref{fig1} (c) and (d) show the spin density obtained for the Weyl-type SOC under the N\'eel and the Bloch domain walls. In the case of the N\'eel wall, only the $z$-component of the spin density was finite, whereas the $y$-component became finite in case of the Bloch wall. The disappearance of the $y$-component in the N\'eel wall occurred due to the contribution of the second term of the last expression in Eq. (\ref{spin torque}). For both magnetic configurations, the numerical results were consistent with the analytical expression obtained by Eq. (\ref{result}), presented in insets of FIG. \ref{fig1}. This result can be understood by the correspondence between magnetization and axial vector potential in the lattice Hamiltonian. As shown above using the continuum model, all components of the magnetization could be regarded as axial vector potentials for the Weyl-type SOC when the Fermi energy was located near the Weyl nodes.

Only the $z$-component of magnetization acted as axial vector potential for the Rashba-type SOC and the $s_z$-conserved SOC. This indicates that the relation between spin density and the axial currents only holds for their $z$-components for the Rashba-type SOC and the $s_z$-conserved type SOC. Indeed, the $z$-component of spin density obtained with the Rashba-type SOC showed a similar spatial profile to that of the Weyl-type SOC, as shown in FIGs. \ref{fig1} (e) and (f). On the contrary, the perpendicular components showed different behaviors due to the absence of the correspondence. The Rashba-type SOC, however, still featured the relation between spin density and axial currents when $J_1=0$. The low-energy effective Hamiltonian for the Rashba-type SOC is given as 
$
H_{\rm RSO} (\bm k)= \hbar v_F\tau_x(k_x\sigma_y-k_y\sigma_x)+\tau_y\hbar v_Fk_z - J_0\hat{\bm M}\cdot \bm \sigma.
$
On the $k_z=0$ plane, the Hamiltonian can be transformed by unitary transformation, ${\rm U} = (\tau_x+\tau_z)/\sqrt{2}$, into
\begin{eqnarray}
\tilde{H}_{\rm RSO} (\bm k)& \equiv& {\rm U}^\dagger H_{\rm RSO} (\bm k){\rm U}
\nonumber =\hbar v_F \tau_z\left[\left(k_x-\tau_z\frac{J_0}{\hbar v_F}M_y\right)\sigma_y-\left(k_y+\tau_z\frac{J_0}{\hbar v_F}M_x\right)\sigma_x\right].
\end{eqnarray}
This suggests that the perpendicular components of magnetization acted as axial vector potential, namely $\bm A_{5\perp} = (J_0/e v_F)\hat{\bm M}\times \hat{\bm z} $, and that there was a correspondence between axial current and spin density, given by $\braket{\bm \sigma}=-\braket{\bm j_{5\perp}}\times\hat{\bm z}/(ev_F)$. The spin density corresponding to the axial Hall current, therefore, can be obtained by
\begin{eqnarray}
\braket{\bm \sigma_{\perp}} \propto \left[\bm \nabla\times(\hat{\bm z}\times \hat{\bm M})\right]\times \bm E.
\label{result_ani}
\end{eqnarray}
Although the one-to-one correspondence was only present for states at $k_z=0$, the relation approximately held when $\sqrt{k_x^2+k_y^2}>>k_z$, and these states contributed to the induced spin density. The numerical results, shown in FIGs. \ref{fig1} (e) and (f), have the same structure as the analytical expression in Eq. (\ref{result_ani}), presented in the inset.

The $z$-component of spin density for the $s_z$-conserved SOC also showed similar behavior to that of the other two models, which was expected due to the correspondence between the $z$-component of the axial current and the spin density. Given these numerical results, we can conclude that the structure of the $z$-component of spin density was qualitatively described by our analytical expression, $\langle \sigma_z(\bm r) \rangle \propto ([\bm \nabla \times \hat{\bm M}(\bm r)]\times \bm E)_z$, even magnetization varies in an atomic scale and irrelevant to the type of spin-orbit coupling, and the spin-transfer torque associated with the induced spin density can be generally expected.

\subsection*{Motion of the domain wall }
Finally, we estimated the velocity of the domain wall driven by the spin torque, given by Eq. (\ref{spin torque}). To describe the dynamics of the domain wall, we employed Thiele's approach \cite{Thiele1973} to map the LLG equation to the equation of motion for the centre coordinates of the domain wall. We considered that magnetic texture varied along the $x$-direction and assumed that it moved at a constant velocity without deformation, namely $\hat{\bm M} = \hat{\bm M} (x - v_{\rm DW}t)$, where $v_{\rm DW}$ is the velocity of the domain wall. Then, the LLG equation Eq. (\ref{llg}) with the spin torque in Eq. (\ref{spin torque}) can be rewritten as
\begin{eqnarray}
\nonumber \gamma_0 \bm B+v_{\rm DW} \left(\hat{\bm M}\times\nabla_x\hat{\bm M}\right)+\alpha v_{\rm DW}\nabla_x\hat{\bm M}  -  \frac{JS}{\hbar\rho_S}\chi_S E_0\left(\nabla_x \hat{\bm M}-\nabla_x M_x\hat{\bm x}\right)=0
\end{eqnarray}
where we applied the electric field along the $x$-direction, $\bm E = E_0\hat{\bm x}$. By taking the inner product of the equation with $\nabla_x \hat{\bm M}$ and integrating over the space, we obtain
\begin{eqnarray}
 - c_0 \alpha v_{\rm DW} + c_1 \frac{JS}{\hbar\rho_S}\chi_S E_0 = f
\end{eqnarray}
where the coefficients are given by
$
c_0 = \int dx (\nabla_x\hat{\bm M})^2,
$
$
c_1 = \int dx\left[ (\nabla_x\hat{\bm M})^2-(\nabla_xM_x)^2\right],
$
and 
$
f =  \int dx \nabla_x \hat{\bm M}\cdot \bm B.
$
When the external magnetic field is absent, the velocity of the domain wall driven by the spin torque is 
$
v_{\rm DW} = \frac{c_1}{c_0}\frac{JS\chi_S}{\alpha \hbar\rho_S} E_0.
$
For magnetization described by $\hat{\bm M} = 1/\cosh[(x-v_{\rm DW}t)/\xi] \hat{\bm n} + \tanh[(x-v_{\rm DW}t)/\xi] \hat{\bm z}$, where $\xi$ is the width of the domain wall, and $\hat{\bm n} = \hat{\bm x}$ for the N\'eel wall and  $\hat{\bm n} = \hat{\bm y}$ for the Bloch wall, the coefficients are  
$
 c_1 = 2/\xi
 $
 for the Bloch wall and 
 $
 c_1 = 4/(3\xi)
 $
 for the N\'eel wall, while $c_0$ is common to the cases as
$
c_0= 2/\xi
$. 
As a consequence, the Bloch wall moves three-and-a-half times faster than the N\'eel wall under the electric field.

With domain wall width $\xi\approx 100$ nm and exchange coupling constant $JS \approx 1$ eV, the averaged magnitude of the emergent axial magnetic field over the domain wall was approximately estimated to be $|\bm B_5| \approx|JS/(ev_F)\xi^{-1}|\sim100$T. When the Fermi energy was located within $\Delta = 2v_F\sqrt{2e|\bm B_5|}$, the system is considered as the quantum Hall regime. For the typical parameters of the half-metallic Weyl semimetal, $\rm Co_3Sn_2S_2$, Gilbert damping constant $\alpha=0.01$, Fermi energy $E_F=0.1$ eV, and electric field $E_x = 10^5$ V/m, the velocity of the Bloch wall was estimated as $v_{\rm DW} \sim 2.8$ km/s, one order of magnitude higher than that of the ferromagnetic nanowire \cite{Meier2007,Hayashi2007}. 


\section*{Discussion}
We provide a brief discussion of the difference in domain wall dynamics between the wall in the Weyl semimetal and that in conventional metals. In the latter, the spin-transfer torque can be classified into adiabatic and non-adiabatic contributions. Adiabatic torque emerges from the exchange of the spin angular momentum between local moments and itinerant electrons, whereas non-adiabatic torque emerges from spin relaxation processes. Owing to a large spin diffusion length, the non-adiabatic spin-transfer torque is considerably smaller than the adiabatic spin-transfer torque in conventional metals. On the contrary, the spin torque in Weyl semimetals with Weyl-type SOT, Eq. (\ref{spin torque}), consists only of the non-adiabatic contribution because of the strong spin-orbit coupling nature of Weyl semimetals. The Hamiltonian, Eq. (\ref{H0}), can be seen as a strong spin-orbit coupled limit that enhances spin relaxation. Therefore, the non-adiabatic spin-transfer torque in Weyl semimetals is expected to be considerably larger than that in conventional metals. The difference in domain wall dynamics might be significant in the weak electric field regime. In a regime where the applied electric field is smaller than the threshold field determined by magnetic anisotropy, the adiabatic spin-transfer torque cannot drive the magnetic domain wall due to the intrinsic pinning effect. Contrary, domain walls under non-adiabatic spin-transfer torque are less affected by the pinning emerging from the magnetic anisotropy. Therefore, the threshold field with non-adiabatic spin-transfer torque is much smaller than that with adiabatic spin-transfer torque. Because of the enhanced non-adiabatic spin-transfer torque in Weyl semimetals, it is possible to drive the domain walls more efficiently than in conventional metals.

The absence of the adiabatic spin-transfer torque in Weyl semimetal with Weyl-type SOT can be explained by the one-to-one correspondence between the axial current and the spin.
In the derivation of the spin-transfer torque, we argued that the Hall conductivity or the off-diagonal part of the magnetoconductivity gives the non-adiabatic contribution to the spin density.
This suggests that the diagonal part of the magnetoconductivity contributes to the adiabatic spin-transfer torque because the adiabatic contributions are perpendicular to the non-adiabatic contributions.
Due to the Onsager reciprocal relation, the diagonal part of the magnetoconductivity is even under sign change of the magnetic field; namely, it can be series expanded with even order of the axial magnetic fields.
By the one-to-one correspondence between the axial current and the spin, the spin density corresponds to the adiabatic spin torque is also expressed by the series of even number of $\bm B_5 \propto \bm \nabla \times \hat{\bm M}$.
The spin-transfer torque is often defined as the spin torque proportional to the first order spatial derivative of the magnetization.
Therefore, within this definition, the adiabatic spin-transfer torque is absent in the Weyl semimetals with the Weyl-type SOC.
Note that the above argument is applicable only for the Weyl-type SOC case retaining the one-to-one correspondence and when dispersion close to Fermi energy is described by the $k$-linear term.
If the dispersion deviates from the linear, there is no one-to-one correspondence between the spin and the axial current, and the other form of the spin-transfer torque including adiabatic contributions is expected. For example, the Rashba-type SOC and the $s_z$-conserved SOC, the one-to-one correspondence is present only between the $z$-component of magnetization and the axial current. That is, other forms of the spin torque might also be expected for perpendicular components. The conventional current-induced spin torque, however, might be negligibly small when the magnetic texture is steep, as the longitudinal conduction of currents is suppressed in the regime. Therefore, the spin torque given by Eq. (\ref{result}), which is related to the localized charge at the domain wall, is expected to be the dominant contribution. Our numerical results containing all contributions to the spin-transfer torques gave consistent results with the analytical results for non-adiabatic spin-transfer torque, presented in insets of FIG. \ref{fig1}. This suggests that our proposed spin torque provided the dominant contribution.

In our analysis, we neglected electron-electron correlations. As our spin torque is closely related to the localized charge at the domain wall, which might be screened once the correlation effect is considered, the spin torque might also be diminished by the screening effect. However, when the Fermi energy is located close to the Weyl points, the Thomas--Fermi screening length can be greater than domain size owing to the low density of states, and the screening effect can be neglected at the nanoscale.

In conclusion, we derived the electrically-induced spin torque in Weyl semimetals with magnetic texture. Based on the spin--axial current correspondence, we calculated the non-equilibrium spin density induced by electric fields by evaluating the axial current density. 
We thus obtained the analytical expression for non-adiabatic spin-transfer torque, and found that the effect can be understood as the Hall effect of axial current. 
To examine our result for the case of the steep magnetic textures and other types of SOC, we numerically calculated the induced spin density for a lattice Hamiltonian with three spin-orbit couplings: the Weyl-type SOC, Rashba-type SOC, and $s_z$-conserved SOC. 
We found that the $z$-component of spin density, which is responsible for the dynamics of the N\'eel and the Bloch domain walls, is well described by the obtained analytical expression even when the magnetization varies in the atomic scale and has the same structure irrelevant of the difference in spin-orbit coupling, although the perpendicular components of spin density depend on spin-orbit coupling. This indicates that the spin-transfer torque associated with the axial Hall effect is generally expected in magnetic Weyl semimetals. We also analyzed the domain wall dynamics driven by spin-transfer torque and estimated that the velocity of the domain wall can be one order of magnitude larger than that of the ferromagnetic nanowire. 
Surprisingly, the spin-transfer torque was independent of impurity-scattering relaxation time when magnetization varied steeply. In this regime, the longitudinal conductivity asymptotically reached zero \cite{Ominato2017,Kobayashi2018}, and therefore dissipation due to Joule heating was suppressed. Due to the suppression of Joule heating, energy efficiency was expected to be much higher than that of conventional metals. Consequently, the domain wall in the Weyl semimetal can be controlled much more efficiently. While a commercial racetrack memory has not yet been developed because of the energy inefficiency arising from Joule heating, magnetic Weyl semimetals can overcome the relevant challenges, and can be a new candidate for racetrack memories that can yield high performance.

\section*{Methods}
To obtain non-equilibrium current density, we used the Boltzmann transport theory. The Boltzmann transport equation for the non-equilibrium distribution function for Weyl electrons with chirality $\lambda$, $f_\lambda(\bm r,\bm k,t)$, is given by
\begin{eqnarray}
\nonumber \frac{\partial f_\lambda(\bm r,\bm k,t)}{\partial t} + \dot{\bm r}_\lambda\cdot \frac{\partial f_\lambda(\bm r,\bm k,t) }{\partial \bm r} + \dot{\bm k}_\lambda\cdot \frac{\partial f_\lambda(\bm r,\bm k,t) }{\partial \bm k} = -\frac{\delta  f_\lambda(\bm r,\bm k,t) }{\tau(\epsilon(\bm k))} \\
\label{bteq}
\end{eqnarray}
in relaxation time approximation. In this paper, we neglected energy dependence on relaxation time, $\tau(\epsilon(\bm k))\approx \tau$, for simplicity. We considered nonmagnetic impurities, $\hat{V}_{\rm imp} = \int d\bm r \psi(\bm r)^{\dagger}V_{\rm imp}(\bm r)\psi(\bm r)$, where $V_{\rm imp}(\bm r) = \sum_Iu(\bm r - \bm R_I)$ with short-range potential $u(\bm r) = u_0\delta(\bm r)$. After taking the Gaussian average of impurity positions, the relaxation time was obtained as $1/\tau = \pi n_i u_0^2 \nu(E_F)$, where $n_i$ is the impurity concentration and $\nu(E_F)$ is the density of states at Fermi energy. We further assumed that the distribution function was close to equilibrium and expanded it with a small deviation from equilibrium as $f_\lambda(\bm r,\bm k,t) = f_{0 \lambda}(\bm r,\bm k,t)+\delta f_\lambda(\bm r,\bm k,t)$, where the equilibrium distribution function was the Fermi--Dirac distribution function. By expanding Eq. (\ref{bteq}) with respect to the $\delta f_\lambda$, the linearized Boltzmann equation was obtained as
\begin{eqnarray}
\frac{\partial\delta f_\lambda}{\partial t}& +& \dot{\bm r}_\lambda\cdot \left[\frac{\partial f_{0 \lambda}}{\partial \epsilon}\left(\frac{\partial \epsilon_\lambda}{\partial \bm r}\right)+\frac{\partial \delta f_\lambda}{\partial \bm r}\right]+ \dot{\bm k_\lambda}\cdot \left[\bm v_\lambda\left(\frac{\partial f_{0 \lambda}}{\partial \epsilon}\right)+\frac{\partial \delta f_\lambda}{\partial \bm k}\right]=-\frac{\delta f_\lambda}{\tau},
\label{LBT}
\end{eqnarray}
where $\bm r_\lambda$ and $\bm k_\lambda$ are the position and the wave number of the wave packets of the Weyl electrons with chirality $\lambda$, respectively. The dynamics of $\bm r_\lambda$ and $\bm k_\lambda$ were determined by the semiclassical equations of motion, $\dot{\bm r_\lambda} = \frac{\partial \epsilon_\lambda}{\hbar \partial \bm k}$ and $\dot{\bm k_\lambda} = - \frac{\partial \epsilon_\lambda}{\hbar\partial \bm r} - \frac{e}{\hbar}\frac{\partial \epsilon_\lambda}{\hbar \partial \bm k} \times\lambda \bm B_5$. Note that we neglected modifications to the semiclassical equation due to the presence of the Berry curvature and orbital magnetization, as they are irrelevant to the spin torques considered in this paper. By combining the semiclassical equation of motion with Eq. (\ref{LBT}), the non-equilibrium distribution function is obtained up to the second order in spatial gradient as
\begin{eqnarray}
\nonumber \delta f(\bm r,\bm k) =\frac{ev_F\tau^2}{1+\left|\frac{ev_F\tau^2 \bm B_{5}(\bm r)}{\hbar k}\right|^2} \left[\delta_{ij}\frac{1}{\tau}\frac{k_i}{|\bm k|}-\epsilon_{ijk}\frac{ev_F}{\hbar }\frac{k_iB_{5}^k}{|\bm k|^2}\right]E_j\frac{\partial f_0}{\partial \epsilon},
\end{eqnarray}
where $\epsilon_{ijk}$ is the completely anti-symmetric tensor. To obtain the form, we introduced an external electric potential to the Hamiltonian, $H=H_{\rm WSM}-e\phi(\bm r)$. The non-equilibrium electric and axial currents are, then, given by
\begin{eqnarray}
\langle j^e_i(\bm r) \rangle&=& -e\sum_{\lambda = \pm 1}  \int \frac{d^3k}{(2\pi)^3}\delta f_\lambda(\bm r,\bm k,t) \dot{\bm r}_\lambda =\delta_{ij} \frac{e^2E_F^2}{3\hbar^3\pi^2v_F}\frac{\tau}{1+(\omega_C\tau)^2}E_j,\\
\langle j^5_i(\bm r) \rangle&=& -e\sum_{\lambda = \pm 1} \lambda \int \frac{d^3k}{(2\pi)^3}\delta f_\lambda(\bm r,\bm k,t) \dot{\bm r}_\lambda =\epsilon_{ijk} \frac{e^3v_F}{3\hbar^3\pi^2} \frac{\tau^2E_F}{1+(\omega_C\tau)^2}B_5^k E_j.
\end{eqnarray}
Note that in the strong field limit, the longitudinal electrical conductivity asymptotically reached zero, whereas the axial Hall conductivity became independent of relaxation time, $\tau$.


\begin{thebibliography}{48}
\urlstyle{rm}
\expandafter\ifx\csname url\endcsname\relax
  \def\url#1{\texttt{#1}}\fi
\expandafter\ifx\csname urlprefix\endcsname\relax\def\urlprefix{URL }\fi
\expandafter\ifx\csname doiprefix\endcsname\relax\def\doiprefix{DOI: }\fi
\providecommand{\bibinfo}[2]{#2}
\providecommand{\eprint}[2][]{\url{#2}}

\bibitem{Zutic2004}
\bibinfo{author}{{\v{Z}}uti{\'{c}}, I.}, \bibinfo{author}{Fabian, J.} \&
  \bibinfo{author}{Sarma, S.~D.}
\newblock \bibinfo{journal}{\bibinfo{title}{{Spintronics: Fundamentals and
  applications}}}.
\newblock {\emph{{Rev. Mod. Phys.}}}
  \textbf{\bibinfo{volume}{76}}, \bibinfo{pages}{323--410},
  \doiprefix\url{10.1103/RevModPhys.76.323} (\bibinfo{year}{2004}).

\bibitem{Slonczewski1996}
\bibinfo{author}{Slonczewski, J.~C.}
\newblock \bibinfo{journal}{\bibinfo{title}{{Current-driven excitation of
  magnetic multilayers}}}.
\newblock {\emph{{J. Magn. Magn. Mater.}}}
  \textbf{\bibinfo{volume}{159}}, \bibinfo{pages}{L1--L7},
  \doiprefix\url{10.1016/0304-8853(96)00062-5} (\bibinfo{year}{1996}).

\bibitem{Berger1996}
\bibinfo{author}{Berger, L.}
\newblock \bibinfo{journal}{\bibinfo{title}{{Emission of spin waves by a
  magnetic multilayer traversed by a current}}}.
\newblock {\emph{{Phys. Rev. B}}} \textbf{\bibinfo{volume}{54}},
  \bibinfo{pages}{9353--9358}, \doiprefix\url{10.1103/PhysRevB.54.9353}
  (\bibinfo{year}{1996}).

\bibitem{Ralph2008}
\bibinfo{author}{Ralph, D.~C.} \& \bibinfo{author}{Stiles, M.~D.}
\newblock \bibinfo{journal}{\bibinfo{title}{{Spin transfer torques}}}.
\newblock {\emph{{J. Magn. Magn. Mater.}}}
  \textbf{\bibinfo{volume}{320}}, \bibinfo{pages}{1190--1216},
  \doiprefix\url{10.1016/j.jmmm.2007.12.019} (\bibinfo{year}{2008}).

\bibitem{Brataas2012}
\bibinfo{author}{Brataas, A.}, \bibinfo{author}{Kent, A.~D.} \&
  \bibinfo{author}{Ohno, H.}
\newblock \bibinfo{journal}{\bibinfo{title}{{Current-induced torques in
  magnetic materials}}}.
\newblock {\emph{{Nat. Mater.}}} \textbf{\bibinfo{volume}{11}},
  \bibinfo{pages}{372--381}, \doiprefix\url{10.1038/nmat3311}
  (\bibinfo{year}{2012}).

\bibitem{Parkin2008}
\bibinfo{author}{Parkin, S. S.~P.}, \bibinfo{author}{Hayashi, M.} \&
  \bibinfo{author}{Thomas, L.}
\newblock \bibinfo{journal}{\bibinfo{title}{{Magnetic domain-wall racetrack
  memory}}}.
\newblock {\emph{{Science}}} \textbf{\bibinfo{volume}{320}},
  \bibinfo{pages}{190--194}, \doiprefix\url{10.1126/science.1145799}
  (\bibinfo{year}{2008}).


\bibitem{Hasan2010}
\bibinfo{author}{Hasan, M.~Z.} \& \bibinfo{author}{Kane, C.~L.}
\newblock \bibinfo{journal}{\bibinfo{title}{{Colloquium: Topological
  insulators}}}.
\newblock {\emph{{Rev. Mod. Phys.}}}
  \textbf{\bibinfo{volume}{82}}, \bibinfo{pages}{3045--3067},
  \doiprefix\url{10.1103/RevModPhys.82.3045} (\bibinfo{year}{2010}).

\bibitem{Qi2011}
\bibinfo{author}{Qi, X.~L.} \& \bibinfo{author}{Zhang, S.~C.}
\newblock \bibinfo{journal}{\bibinfo{title}{{Topological insulators and
  superconductors}}}.
\newblock {\emph{{Rev. Mod. Phys.}}}
  \textbf{\bibinfo{volume}{83}}, \bibinfo{pages}{1057--1110},
  \doiprefix\url{10.1103/RevModPhys.83.1057} (\bibinfo{year}{2011}).

\bibitem{Nomura2010}
\bibinfo{author}{Nomura, K.} \& \bibinfo{author}{Nagaosa, N.}
\newblock \bibinfo{journal}{\bibinfo{title}{{Electric charging of magnetic
  textures on the surface of a topological insulator}}}.
\newblock {\emph{{Phys. Rev. B}}} \textbf{\bibinfo{volume}{82}},
  \bibinfo{pages}{161401}, \doiprefix\url{10.1103/PhysRevB.82.161401}
  (\bibinfo{year}{2010}).

\bibitem{Hurst2015}
\bibinfo{author}{Hurst, H.~M.}, \bibinfo{author}{Efimkin, D.~K.},
  \bibinfo{author}{Zang, J.} \& \bibinfo{author}{Galitski, V.}
\newblock \bibinfo{journal}{\bibinfo{title}{{Charged skyrmions on the surface
  of a topological insulator}}}.
\newblock {\emph{{Phys. Rev. B}}} \textbf{\bibinfo{volume}{91}},
  \bibinfo{pages}{060401}, \doiprefix\url{10.1103/PhysRevB.91.060401}
  (\bibinfo{year}{2015}).

\bibitem{Garate2010}
\bibinfo{author}{Garate, I.} \& \bibinfo{author}{Franz, M.}
\newblock \bibinfo{journal}{\bibinfo{title}{{Inverse spin-galvanic effect in
  the interface between a topological insulator and a ferromagnet}}}.
\newblock {\emph{{Phys. Rev. Lett.}}}
  \textbf{\bibinfo{volume}{104}}, \bibinfo{pages}{146802},
  \doiprefix\url{10.1103/PhysRevLett.104.146802} (\bibinfo{year}{2010}).

\bibitem{Yokoyama2010}
\bibinfo{author}{Yokoyama, T.}, \bibinfo{author}{Zang, J.} \&
  \bibinfo{author}{Nagaosa, N.}
\newblock \bibinfo{journal}{\bibinfo{title}{{Theoretical study of the dynamics
  of magnetization on the topological surface}}}.
\newblock {\emph{{Phys. Rev. B}}} \textbf{\bibinfo{volume}{81}},
  \bibinfo{pages}{241410}, \doiprefix\url{10.1103/PhysRevB.81.241410}
  (\bibinfo{year}{2010}).

\bibitem{Sakai2014}
\bibinfo{author}{Sakai, A.} \& \bibinfo{author}{Kohno, H.}
\newblock \bibinfo{journal}{\bibinfo{title}{{Spin torques and charge transport
  on the surface of topological insulator}}}.
\newblock {\emph{{Phys. Rev. B}}} \textbf{\bibinfo{volume}{89}},
  \bibinfo{pages}{165307}, \doiprefix\url{10.1103/PhysRevB.89.165307}
  (\bibinfo{year}{2014}).

\bibitem{Fan2014}
\bibinfo{author}{Fan, Y.} \emph{et~al.}
\newblock \bibinfo{journal}{\bibinfo{title}{{Magnetisation switching through
  giant spin-orbit torque in a magnetically doped topological insulator heterostructure}}}.
\newblock {\emph{{Nat. Mater.}}} \textbf{\bibinfo{volume}{13}},
  \bibinfo{pages}{699--704}, \doiprefix\url{10.1038/nmat3973}
  (\bibinfo{year}{2014}).

\bibitem{Shiomi2014}
\bibinfo{author}{Shiomi, Y.} \emph{et~al.}
\newblock \bibinfo{journal}{\bibinfo{title}{{Spin-electricity conversion
  induced by spin injection into topological insulators}}}.
\newblock {\emph{{Phys. Rev. Lett.}}}
  \textbf{\bibinfo{volume}{113}}, \bibinfo{pages}{196601},
  \doiprefix\url{10.1103/PhysRevLett.113.196601} (\bibinfo{year}{2014}).

\bibitem{Kondou2016}
\bibinfo{author}{Kondou, K.} \emph{et~al.}
\newblock \bibinfo{journal}{\bibinfo{title}{{Fermi-level-dependent
  charge-to-spin current conversion by Dirac surface states of
  topological insulators}}}.
\newblock {\emph{{Nat. Phys.}}} \textbf{\bibinfo{volume}{12}},
  \bibinfo{pages}{1027--1031}, \doiprefix\url{10.1038/nphys3833}
  (\bibinfo{year}{2016}).

\bibitem{Okuma2016}
\bibinfo{author}{Okuma, N.} \& \bibinfo{author}{Nomura, K.}
\newblock \bibinfo{journal}{\bibinfo{title}{{Microscopic derivation of magnon
  spin current in a topological insulator/ferromagnet heterostructure}}}.
\newblock {\emph{{Phys. Rev. B}}} \textbf{\bibinfo{volume}{95}},
  \bibinfo{pages}{1--9}, \doiprefix\url{10.1103/PhysRevB.95.115403}
  (\bibinfo{year}{2017}).

\bibitem{ShuichiMurakami2007}
\bibinfo{author}{Murakami, S.}
\newblock \bibinfo{journal}{\bibinfo{title}{{Phase transition between the
  quantum spin Hall and insulator phases in 3D: Emergence of a topological
  gapless phase}}}.
\newblock {\emph{{New J. Phys.}}} \textbf{\bibinfo{volume}{9}},
  \bibinfo{pages}{356--356}, \doiprefix\url{10.1088/1367-2630/9/9/356}
  (\bibinfo{year}{2007}).

\bibitem{Wan2011}
\bibinfo{author}{Wan, X.}, \bibinfo{author}{Turner, A.~M.},
  \bibinfo{author}{Vishwanath, A.} \& \bibinfo{author}{Savrasov, S.~Y.}
\newblock \bibinfo{journal}{\bibinfo{title}{{Topological semimetal and
  Fermi-arc surface states in the electronic structure of pyrochlore
  iridates}}}.
\newblock {\emph{{Phys. Rev. B}}} \textbf{\bibinfo{volume}{83}},
  \bibinfo{pages}{205101}, \doiprefix\url{10.1103/PhysRevB.83.205101}
  (\bibinfo{year}{2011}).

\bibitem{Burkov2011}
\bibinfo{author}{Burkov, A.~A.} \& \bibinfo{author}{Balents, L.}
\newblock \bibinfo{journal}{\bibinfo{title}{{Weyl semimetal in a topological
  insulator multilayer}}}.
\newblock {\emph{{Phys. Rev. Lett.}}}
  \textbf{\bibinfo{volume}{107}}, \bibinfo{pages}{127205},
  \doiprefix\url{10.1103/PhysRevLett.107.127205} (\bibinfo{year}{2011}).

\bibitem{Weyl1929}
\bibinfo{author}{Weyl, H.}
\newblock \bibinfo{journal}{\bibinfo{title}{{Elektron und Gravitation. I}}}.
\newblock {\emph{{Zeitschrift f{\"{u}}r Phys.}}}
  \textbf{\bibinfo{volume}{56}}, \bibinfo{pages}{330--352},
  \doiprefix\url{10.1007/BF01339504} (\bibinfo{year}{1929}).

\bibitem{Nielsen1981}
\bibinfo{author}{Nielsen, H.~B.} \& \bibinfo{author}{Ninomiya, M.}
\newblock \bibinfo{journal}{\bibinfo{title}{{Absence of neutrinos on a lattice.
  (I). Proof by homotopy theory}}}.
\newblock {\emph{{Nucl. Physics, Sect. B}}}
  \textbf{\bibinfo{volume}{185}}, \bibinfo{pages}{20--40},
  \doiprefix\url{10.1016/0550-3213(81)90361-8} (\bibinfo{year}{1981}).

\bibitem{Hosur2013}
\bibinfo{author}{Hosur, P.} \& \bibinfo{author}{Qi, X.}
\newblock \bibinfo{journal}{\bibinfo{title}{{Recent developments in transport
  phenomena in Weyl semimetals}}}.
\newblock {\emph{{Comptes Rendus Phys.}}}
  \textbf{\bibinfo{volume}{14}}, \bibinfo{pages}{857--870},
  \doiprefix\url{10.1016/j.crhy.2013.10.010} (\bibinfo{year}{2013}).

\bibitem{Bulmash2014}
\bibinfo{author}{Bulmash, D.}, \bibinfo{author}{Liu, C.~X.} \&
  \bibinfo{author}{Qi, X.~L.}
\newblock \bibinfo{journal}{\bibinfo{title}{{Prediction of a Weyl semimetal in
  Hg$_{1-x-y}$Cd$_x$Mn$_y$Te}}}.
\newblock {\emph{{Phys. Rev. B}}} \textbf{\bibinfo{volume}{89}},
  \bibinfo{pages}{081106}, \doiprefix\url{10.1103/PhysRevB.89.081106}
  (\bibinfo{year}{2014}).

\bibitem{Kurebayashi2014}
\bibinfo{author}{Kurebayashi, D.} \& \bibinfo{author}{Nomura, K.}
\newblock \bibinfo{journal}{\bibinfo{title}{{Weyl semimetal phase in
  solid-solution narrow-gap semiconductors}}}.
\newblock {\emph{{J. Phys. Soc. Japan}}}
  \textbf{\bibinfo{volume}{83}}, \bibinfo{pages}{063709},
  \doiprefix\url{10.7566/JPSJ.83.063709} (\bibinfo{year}{2014}).

\bibitem{Xu2017a}
\bibinfo{author}{Xu, Q.} \emph{et~al.}
\newblock \bibinfo{journal}{\bibinfo{title}{{Topological surface Fermi arcs in
  the magnetic Weyl semimetal Co$_3$Sn$_2$S$_2$}}}.
\newblock {\emph{{Phys. Rev. B}}} \textbf{\bibinfo{volume}{97}},
  \bibinfo{pages}{235416}, \doiprefix\url{10.1103/PhysRevB.97.235416}
  (\bibinfo{year}{2018}).

\bibitem{Liu2018}
\bibinfo{author}{Liu, E.} \emph{et~al.}
\newblock \bibinfo{journal}{\bibinfo{title}{{Giant anomalous Hall effect in a
  ferromagnetic kagome-lattice semimetal}}}.
\newblock {\emph{{Nat. Phys.}}}
  \doiprefix\url{10.1038/s41567-018-0234-5} (\bibinfo{year}{2018}).

\bibitem{Araki2016a}
\bibinfo{author}{Araki, Y.}, \bibinfo{author}{Yoshida, A.} \&
  \bibinfo{author}{Nomura, K.}
\newblock \bibinfo{journal}{\bibinfo{title}{{Universal charge and current on
  magnetic domain walls in Weyl semimetals}}}.
\newblock {\emph{{Phys. Rev. B}}} \textbf{\bibinfo{volume}{94}},
  \bibinfo{pages}{1--7}, \doiprefix\url{10.1103/PhysRevB.94.115312}
  (\bibinfo{year}{2016}).

\bibitem{Araki2018a}
\bibinfo{author}{Araki, Y.}, \bibinfo{author}{Yoshida, A.} \&
  \bibinfo{author}{Nomura, K.}
\newblock \bibinfo{journal}{\bibinfo{title}{{Localized charge in various
  configurations of magnetic domain wall in a Weyl semimetal}}}.
\newblock {\emph{{Phys. Rev. B}}} \textbf{\bibinfo{volume}{98}},
  \bibinfo{pages}{1--10}, \doiprefix\url{10.1103/PhysRevB.98.045302}
  (\bibinfo{year}{2018}).

\bibitem{Zhang2004}
\bibinfo{author}{Zhang, S} \& \bibinfo{author}{Li, Z.}
\newblock \bibinfo{journal}{\bibinfo{title}{{
Roles of nonequilibrium conduction electrons on the magnetization dynamics of ferromagnets
}}}.
\newblock {\emph{{Phys. Rev. Lett.}}} \textbf{\bibinfo{volume}{93}},
  \bibinfo{pages}{1-4}, \doiprefix\url{10.1103/PhysRevLett.93.127204}
  (\bibinfo{year}{2004}).


\bibitem{Kohno2006}
\bibinfo{author}{Kohno, H.}, \bibinfo{author}{Tatara, G.} \&
  \bibinfo{author}{Shibata, J.}
\newblock \bibinfo{journal}{\bibinfo{title}{{Microscopic calculation of spin
  torques in disordered ferromagnets}}}.
\newblock {\emph{{J. Phys. Soc. Jpn.}}}
  \textbf{\bibinfo{volume}{75}}, \bibinfo{pages}{113706},
  \doiprefix\url{10.1143/JPSJ.75.113706} (\bibinfo{year}{2006}).


\bibitem{Hals2015}
\bibinfo{author}{Hals, K. M. D.} \& \bibinfo{author}{Brataas, A.}
\newblock \bibinfo{journal}{\bibinfo{title}{{Spin-motive forces and current-induced torques in ferromagnets}}}.
\newblock {\emph{{Phys. Rev. B}}}
\textbf{\bibinfo{volume}{91}}, \bibinfo{pages}{214401},
\doiprefix\url{10.1103/PhysRevB.91.214401} (\bibinfo{year}{2015}).

\bibitem{Eltschka2010}
\bibinfo{author}{Eltschka, M.} \emph{et~al.}
\newblock \bibinfo{journal}{\bibinfo{title}{{Nonadiabatic Spin Torque Investigated Using Thermally Activated Magnetic Domain Wall Dynamics}}}.
\newblock {\emph{{Phys. Rev. Lett.}}}
\textbf{\bibinfo{volume}{105}}, \bibinfo{pages}{056601},
\doiprefix\url{10.1103/PhysRevLett.105.05660} (\bibinfo{year}{2010}).

\bibitem{Zyuzin2012a}
\bibinfo{author}{Zyuzin, A.~A.} \& \bibinfo{author}{Burkov, A.~A.}
\newblock \bibinfo{journal}{\bibinfo{title}{{Topological response in Weyl
  semimetals and the chiral anomaly}}}.
\newblock {\emph{{Phys. Rev. B}}} \textbf{\bibinfo{volume}{86}},
  \bibinfo{pages}{115133}, \doiprefix\url{10.1103/PhysRevB.86.115133}
  (\bibinfo{year}{2012}).

\bibitem{Araki2018}
\bibinfo{author}{Araki, Y.} \& \bibinfo{author}{Nomura, K.}
\newblock \bibinfo{journal}{\bibinfo{title}{{Charge pumping induced by magnetic
  texture dynamics in Weyl semimetals}}}.
\newblock {\emph{{Phys. Rev. Appl.}}}
  \textbf{\bibinfo{volume}{10}}, \bibinfo{pages}{1},
  \doiprefix\url{10.1103/PhysRevApplied.10.014007} (\bibinfo{year}{2018}).

\bibitem{Zhang2009}
\bibinfo{author}{Zhang, H.} \emph{et~al.}
\newblock \bibinfo{journal}{\bibinfo{title}{{Topological insulators in Bi$_2$Se$_3$,
  Bi$_2$Te$_3$ and Sb$_2$Te$_3$ with a single Dirac cone on the surface}}}.
\newblock {\emph{{Nat. Phys.}}} \textbf{\bibinfo{volume}{5}},
  \bibinfo{pages}{438--442}, \doiprefix\url{10.1038/nphys1270}
  (\bibinfo{year}{2009}).

\bibitem{Liu2010}
\bibinfo{author}{Liu, C.~X.} \emph{et~al.}
\newblock \bibinfo{journal}{\bibinfo{title}{{Model Hamiltonian for topological
  insulators}}}.
\newblock {\emph{{Phys. Rev. B}}} \textbf{\bibinfo{volume}{82}},
  \bibinfo{pages}{045122}, \doiprefix\url{10.1103/PhysRevB.82.045122}
  (\bibinfo{year}{2010}).

\bibitem{Wang2012}
\bibinfo{author}{Wang, Z.} \emph{et~al.}
\newblock \bibinfo{journal}{\bibinfo{title}{{Dirac semimetal and topological
  phase transitions in A$_3$Bi (A=Na, K, Rb)}}}.
\newblock {\emph{{Phys. Rev. B}}} \textbf{\bibinfo{volume}{85}},
  \bibinfo{pages}{195320}, \doiprefix\url{10.1103/PhysRevB.85.195320}
  (\bibinfo{year}{2012}).

\bibitem{Morimoto2014}
\bibinfo{author}{Morimoto, T.} \& \bibinfo{author}{Furusaki, A.}
\newblock \bibinfo{journal}{\bibinfo{title}{{Weyl and Dirac semimetals with Z2
  topological charge}}}.
\newblock {\emph{{Phys. Rev. B}}} \textbf{\bibinfo{volume}{89}},
  \bibinfo{pages}{1--14}, \doiprefix\url{10.1103/PhysRevB.89.235127}
  (\bibinfo{year}{2014}).

\bibitem{Wang2013a}
\bibinfo{author}{Wang, Z.}, \bibinfo{author}{Weng, H.}, \bibinfo{author}{Wu,
  Q.}, \bibinfo{author}{Dai, X.} \& \bibinfo{author}{Fang, Z.}
\newblock \bibinfo{journal}{\bibinfo{title}{{Three-dimensional Dirac semimetal
  and quantum transport in Cd3As2}}}.
\newblock {\emph{{Phys. Rev. B}}} \textbf{\bibinfo{volume}{88}},
  \bibinfo{pages}{125427}, \doiprefix\url{10.1103/PhysRevB.88.125427}
  (\bibinfo{year}{2013}).

\bibitem{Liu2014}
\bibinfo{author}{Liu, Z.~K.} \emph{et~al.}
\newblock \bibinfo{journal}{\bibinfo{title}{{Discovery of a three-dimensional
  topological Dirac semimetal, Na$_3$Bi}}}.
\newblock {\emph{{Science.}}} \textbf{\bibinfo{volume}{343}},
  \bibinfo{pages}{864--867}, \doiprefix\url{10.1126/science.1245085}
  (\bibinfo{year}{2014}).

\bibitem{Liu2014b}
\bibinfo{author}{Liu, Z.~K.} \emph{et~al.}
\newblock \bibinfo{journal}{\bibinfo{title}{{A stable three-dimensional
  topological Dirac semimetal $\rm Cd_3As_2$}}}.
\newblock {\emph{{Nat. Mater.}}} \textbf{\bibinfo{volume}{13}},
  \bibinfo{pages}{677--681}, \doiprefix\url{10.1038/nmat3990}
  (\bibinfo{year}{2014}).

\bibitem{Ozawa2018}
Ozawa, A. \& Nomura, K. (Private communication, 2018).

\bibitem{Burkov2018}
\bibinfo{author}{Burkov, A.~A.}
\newblock \bibinfo{journal}{\bibinfo{title}{{Quantum anomalies in nodal line
  semimetals}}}.
\newblock {\emph{{Phys. Rev. B}}} \textbf{\bibinfo{volume}{97}},
  \bibinfo{pages}{1--9}, \doiprefix\url{10.1103/PhysRevB.97.165104}
  (\bibinfo{year}{2018}).

\bibitem{Thiele1973}
\bibinfo{author}{Thiele, A.~A.}
\newblock \bibinfo{journal}{\bibinfo{title}{{Steady-state motion of magnetic
  domains}}}.
\newblock {\emph{{Phys. Rev. Lett.}}}
  \textbf{\bibinfo{volume}{30}}, \bibinfo{pages}{230--233},
  \doiprefix\url{10.1103/PhysRevLett.30.230} (\bibinfo{year}{1973}).

\bibitem{Meier2007}
\bibinfo{author}{Meier, G.} \emph{et~al.}
\newblock \bibinfo{journal}{\bibinfo{title}{{Direct imaging of stochastic
  domain-wall motion driven by nanosecond current pulses}}}.
\newblock {\emph{{Phys. Rev. Lett.}}}
  \textbf{\bibinfo{volume}{98}}, \bibinfo{pages}{1--4},
  \doiprefix\url{10.1103/PhysRevLett.98.187202} (\bibinfo{year}{2007}).

\bibitem{Hayashi2007}
\bibinfo{author}{Hayashi, M.} \emph{et~al.}
\newblock \bibinfo{journal}{\bibinfo{title}{{Current driven domain wall
  velocities exceeding the spin angular momentum transfer rate in permalloy
  nanowires}}}.
\newblock {\emph{{Phys. Rev. Lett.}}}
  \textbf{\bibinfo{volume}{98}}, \bibinfo{pages}{1--4},
  \doiprefix\url{10.1103/PhysRevLett.98.037204} (\bibinfo{year}{2007}).

\bibitem{Ominato2017}
\bibinfo{author}{Ominato, Y.}, \bibinfo{author}{Kobayashi, K.} \&
  \bibinfo{author}{Nomura, K.}
\newblock \bibinfo{journal}{\bibinfo{title}{{Anisotropic magnetotransport in
  Dirac--Weyl magnetic junctions}}}.
\newblock {\emph{{Phys. Rev. B}}} \textbf{\bibinfo{volume}{95}},
  \bibinfo{pages}{1--6}, \doiprefix\url{10.1103/PhysRevB.95.085308}
  (\bibinfo{year}{2017}).

\bibitem{Kobayashi2018}
\bibinfo{author}{Kobayashi, K.}, \bibinfo{author}{Ominato, Y.} \&
  \bibinfo{author}{Nomura, K.}
\newblock \bibinfo{journal}{\bibinfo{title}{{Helicity-protected domain-wall
  magnetoresistance in ferromagnetic Weyl semimetal}}}.
\newblock {\emph{{J. Phys. Soc. Japan}}}
  \textbf{\bibinfo{volume}{87}}, \bibinfo{pages}{1--5},
  \doiprefix\url{10.7566/JPSJ.87.073707} (\bibinfo{year}{2018}).

\end{thebibliography}


 
\section*{Acknowledgements}
The authors are grateful to H. Kohno and Y. Araki for insightful discussions.
This work was supported by the KAKENHI Grant No. JP17K05485, a Grant-in-Aid for Scientific Research on Innovative Areas "Topological Materials Science" (KAKENHI Grant No. JP15H05854), from the Japan Society for the Promotion of Science, and JST CREST (JPMJCR18T2).
D. K. was supported by a JPSJ Research Fellowship for Young Scientists and the RIKEN Special Postdoctoral Researcher Program.

\section*{Author contributions }
D.K. performed the numerical calculations. Both D.K. and K.N. contributed to the theoretical framework and the interpretation of the results, as well as to the preparation of this manuscript.

\section{Additional information}
{\bf Competing interests:} The authors declare no competing financial interests.

\end{document}